\documentclass[letter]{aa} 

\usepackage{graphicx}
\usepackage{txfonts}

\begin{document}
   \title{Companion stars of Type Ia supernovae and hypervelocity stars}

   \author{Bo Wang
          \inst{1,2}
          \and
          Zhanwen Han \inst{1}
          }


   \institute{National Astronomical Observatories/Yunnan Observatory,
              the Chinese Academy of Sciences, Kunming 650011, China\\
              \email{wangbo@ynao.ac.cn}
              \and
              Graduate University of the Chinese Academy of Sciences, Beijing 100049, China}

   \date{Received ; accepted}


  \abstract
{Recent investigations of the white dwarf (WD) + He star channel of
Type Ia supernovae (SNe Ia) imply that this channel can produce SNe
Ia with short delay times. The companion stars in this channel would
survive and potentially be identifiable.}
{In this {\em Letter}, we study the properties of the companion
stars of this channel at the moment of SN explosion, which can be
verified by future observations.}
{According to SN Ia production regions of the WD + He star channel
and three formation channels of WD + He star systems, we performed a
detailed binary population synthesis study to obtain the properties
of the surviving companions.}
{We obtained the distributions of many properties of the companion
stars of this channel at the moment of SN explosion. We find that
the surviving companion stars have a high spatial velocity
($>$400\,km/s) after SN explosion, which could be an alternative
origin for hypervelocity stars (HVSs), especially for HVSs such as
US 708.}
   {}


\titlerunning{Companion stars of SNe Ia and HVSs}
\authorrunning{B. Wang \& Z. Han}

   \maketitle

%

\section{Introduction} \label{1. Introduction}
Type Ia supernovae (SNe Ia) play an important role in the study of
cosmic evolution, especially in cosmology. They have been applied
successfully in determining cosmological parameters (e.g., $\Omega$
and $\Lambda$; Riess et al. 1998; Perlmutter et al. 1999). It is
generally believed that SNe Ia are thermonuclear explosions of
carbon--oxygen white dwarfs (CO WDs) in binaries (for the review see
Nomoto et al. 1997). However, there is still no agreement on the
nature of their progenitors (Hillebrandt \& Niemeyer 2000;
Podsiadlowski et al. 2008; Wang et al. 2008), and no SN Ia
progenitor system has been conclusively identified from before the
explosion.

Over the past few decades, two families of SN Ia progenitor models
have been proposed, i.e., the double-degenerate (DD) and
single-degenerate (SD) models. Of these two models, the SD model is
widely accepted at present (Nomoto et al. 1984). It is suggested
that the DD model, which involves the merger of two CO WDs (Iben \&
Tutukov 1984; Webbink 1984; Han 1998), likely leads to an
accretion-induced collapse rather than to an SN Ia (Nomoto \& Iben
1985). For the SD model, the companion is probably a main-sequence
(MS) star, a slightly evolved subgiant star (WD + MS channel), or a
red-giant star (WD + RG channel) (e.g., Hachisu et al. 1996,
1999a,b; Li $\&$ van den Heuvel 1997; Langer et al. 2000; Han $\&$
Podsiadlowski 2004, 2006; Chen $\&$ Li 2007, 2009; Meng et al. 2009;
L\"{u} et al. 2009; Wang, Li \& Han 2009). An explosion following
the merger of two WDs would leave no remnant, while the companion
star in the SD model would survive and be potentially identifiable
(Podsiadlowski 2003). There has been no conclusive proof yet that
any individual object is the surviving companion star of an SN Ia.
It will be a promising method to test SN Ia progenitor models by
identifying their surviving companion stars.

Yoon $\&$ Langer (2003) followed the evolution of a CO WD + He star
system with a $1.0\,M_{\odot}$ CO WD and a $1.6\,M_{\odot}$ He star
in a 0.124\,d orbit. In this binary, the WD accretes He from the He
star and grows in mass to the Chandrasekhar (Ch) mass. SNe Ia from
this binary channel can neatly avoid H lines. Recently, Wang et al.
(2009a) systematically studied the WD + He star channel of SNe Ia.
In the study, they carried out binary evolution calculations of this
channel for about 2600 close WD binaries, in which a CO WD accretes
material from an He MS star or an He subgiant to increase its mass
to the Ch mass. The study shows the parameter spaces for the
progenitors of SNe Ia. By using a detailed binary population
synthesis (BPS) approach, Wang et al. (2009b) find that the Galactic
SN Ia birthrate from this channel is $\sim$$0.3\times 10^{-3}\ {\rm
yr}^{-1}$ and that this channel can produce SNe Ia with short delay
times ($\sim$45$-$140\,Myr) from the star formation to SN explosion.
The companion star in this channel would survive and show
distinguishing properties.

In recent years hypervelocity stars (HVSs) have been observed in the
halo of the Galaxy. HVSs are stars with a velocity so great that
they are able to escape the gravitational pull of the Galaxy.
However, the formation of HVSs is still unclear (for a recent review
see Tutukov \& Fedorova 2009). It has been suggested that such HVSs
can be formed by the tidal disruption of a binary through
interaction with the super-massive black hole (SMBH) at the Galactic
center (GC) (Hills 1988; Yu \& Tremaine 2003). The first three HVSs
have only recently been discovered serendipitously (e.g., Brown et
al. 2005; Hirsch et al. 2005; Edelmann et al. 2005). Up to now,
about 17 HVSs have been discovered in the Galaxy (Brown et al. 2009;
Tillich et al. 2009), most of which are B-type stars, probably with
masses ranging from 3 to 5\,$M_\odot$ (Brown et al. 2005, 2009;
Edelmann et al. 2005). One HVS, HE 0437-5439, is known to be an
apparently normal early B-type star. Edelmann et al. (2005) suggests
that the star could have originated in the Large Magellanic Cloud,
because it is much closer to this galaxy (18 kpc) than to the GC
(see also Przybilla et al. 2008).

At present, only one HVS, US 708, is a subdwarf O (sdO) star, and
Hirsch et al. (2005) speculatS that US 708 is formed by the merger
of two He WDs in a close binary induced by the interaction with the
SMBH in the GC and then escaped. Recently, Perets (2009) has
suggested that US 708 may have been ejected as a binary from a
triple disruption by the SMBH, which later on evolved and merged to
form an sdO star. Because of the existence of the short orbital
periods ($\sim$1\,h) for the WD + He star systems, Justham et al.
(2009) argues that the WD + He star channel of SNe Ia may provide a
natural explanation for stars like US 708.

Han (2008a) obtained the distributions of many properties of the
surviving companions from the WD + MS channel of SNe Ia. The
properties can be verified by future observations. The purpose of
this {\em Letter} is to investigate the properties of the surviving
companions of the WD + He star channel and to explore whether HVSs
such as US 708 could have been released from the binaries that
produced SNe Ia. In Section 2, we describe the BPS approach and the
simulation results for the properties of the surviving companions.
Finally, a discussion is given in Section 3.

\section{The distributions of properties of the companion stars} \label{2. The distributions of properties of the companion stars}

In the WD + He star channel, the progenitor of an SN Ia is a close
WD binary system, which has most likely emerged from the CE
evolution (Paczy\'{n}ski 1976) of a giant binary system. The CE
ejection is still an open problem. Here, we use the standard energy
equations (Webbink 1984) to calculate the output of the CE phase.
The CE is ejected if
\begin{equation}
 \alpha_{\rm ce} \left( {G M_{\rm don}^{\rm f} M_{\rm acc} \over 2 a_{\rm f}}
- {G M_{\rm don}^{\rm i} M_{\rm acc} \over 2 a_{\rm i}} \right) = {G
M_{\rm don}^{\rm i} M_{\rm env} \over \lambda R_{\rm don}},
\end{equation}
where $\lambda$ is a structure parameter that depends on the
evolutionary stage of the donor, $M_{\rm don}$ is the mass of the
donor, $M_{\rm acc}$ the mass of the accretor, $a$ the orbital
separation, $M_{\rm env}$ the mass of the donor's envelope, $R_{\rm
don}$ the radius of the donor, and the indices ${\rm i}$ and ${\rm
f}$ denote the initial and final values, respectively. The right
side of the equation represents the binding energy of the CE, the
left side shows the difference between the final and initial orbital
energy, and $\alpha_{\rm ce}$ is the CE ejection efficiency, i.e.,
the fraction of the released orbital energy used to eject the CE.
For this prescription of the CE ejection, there are two highly
uncertain parameters (i.e., $\lambda$ and $\alpha_{\rm ce}$). As in
previous studies, we combine $\alpha_{\rm ce}$ and $\lambda$ into
one free parameter $\alpha_{\rm ce}\lambda$, and set it to be 0.5
(e.g., Wang et al. 2009b).

To obtain the distributions of properties of the surviving
companions, we performed a Monte Carlo simulation in the BPS study.
In the simulation, by using the Hurley's rapid binary evolution code
(Hurley et al. 2000, 2002), we followed the evolution of
$4\times10^{\rm 7}$ sample binaries from the star formation to the
formation of the WD + He star systems according to three
evolutionary channels (i.e., the He star channel, the EAGB channel,
and the TPAGB channel; for details see Wang et al. 2009b). If a
binary system evolves to a WD + He star system, and if the system,
at the beginning of the Roche lobe overflow (RLOF) phase, is located
in the SN Ia production regions in the plane of ($\log P^{\rm i}$,
$M_2^{\rm i}$) for its $M_{\rm WD}^{\rm i}$, where $P^{\rm i}$,
$M_2^{\rm i}$ and $M_{\rm WD}^{\rm i}$ are, respectively, the
orbital period, the secondary's mass, and the WD's mass of the WD +
He star system at the beginning of the RLOF (see Fig. 8 of Wang et
al. 2009a), we assume that an SN Ia is produced, and the properties
of the WD binary at the moment of SN explosion are obtained by
interpolation in the three-dimensional grid ($M_{\rm WD}^{\rm i}$,
$M_2^{\rm i}$, $\log P^{\rm i}$) of the $\sim$2600 close WD binaries
calculated in Wang et al. (2009a).

In the BPS study, the primordial binary samples are generated in the
Monte Carlo way. We adopted the following input for the simulation
(e.g., Han et al. 1995a, 2002, 2003, 2007; Han 2008b; Wang et al.
2009b). (1) The star-formation rate (SFR) is taken to be constant
over the past 15\,Gyr. For the constant SFR, we assume that a binary
with its primary more massive than $0.8\,M_{\odot}$ is formed
annually (e.g., Han et al. 1995b). (2) The initial mass function
(IMF) of Miller \& Scalo (1979) is adopted. (3) The mass-ratio
distribution is taken to be constant. (4) The distribution of
separations is taken to be constant in $\log a$ for wide binaries,
where $a$ is the orbital separation. (5) The orbits are assumed to
be circular.

The simulation gives current-epoch distributions of many properties
of companions at the moment of SN explosion, e.g., the masses, the
orbital periods, the orbital separations, the orbital velocities,
the effective temperatures, the luminosities, the surface gravities,
the surface abundances, the mass-transfer rates, the mass-loss rates
of the optically thick stellar winds, etc. The simulation also shows
the initial parameters of the primordial binaries and the WD
binaries that lead to SNe Ia. Figures 1-5 are selected distributions
that may be helpful for identifying the surviving companion stars.

\section{Discussion}\label{3. Discussion}
\begin{figure}
\includegraphics[width=5.6cm,angle=270]{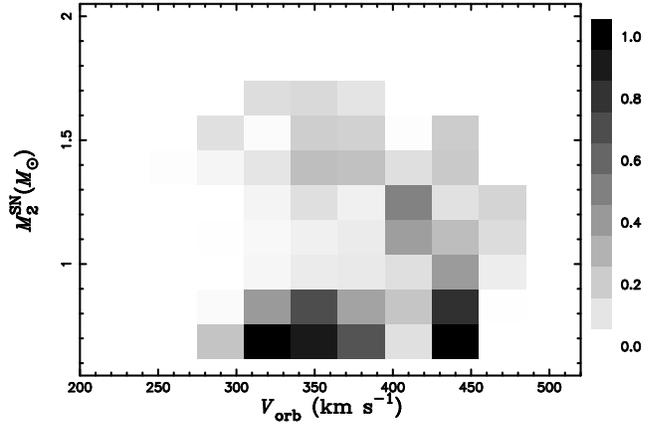}
\caption{The distribution of properties of companion stars in the
plane of ($V_{\rm orb}$, $M_2^{\rm SN}$) at the current epoch, where
$V_{\rm orb}$ is the orbital velocity  and $M_2^{\rm SN}$ the mass
at the moment of SN explosion.}
\end{figure}

\begin{figure}
\includegraphics[width=5.6cm,angle=270]{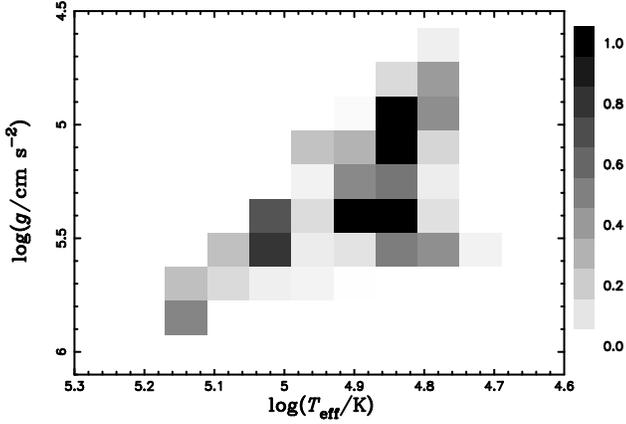}
\caption{Similar to Fig. 1, but in the plane of ($\log T_{\rm eff}$,
$\log g$), where $T_{\rm eff}$ is the effective temperature of
companion stars at the moment  of SN explosion and $\log g$ the
surface gravity. The companion stars are out of thermal equilibrium,
e.g., a 1.243$\,M_\odot$ He star with 0.274$\,R_\odot$ and ($\log
T_{\rm eff}$, $\log g$)=(4.70, 5.65) will be a 0.267$\,R_\odot$ He
star and with ($\log T_{\rm eff}$, $\log g$)=(4.75, 5.67) after the
He star back to the thermal equilibrium.}
\end{figure}

\begin{figure}
\includegraphics[width=8.6cm,angle=0]{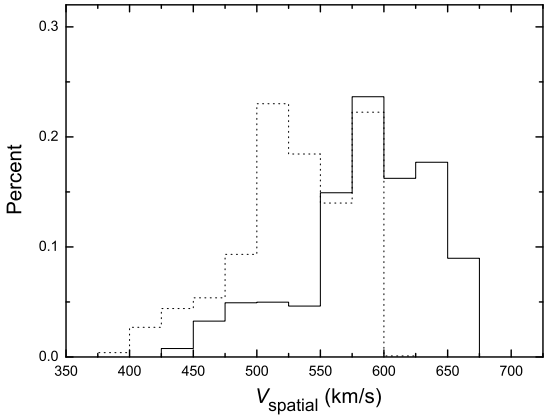}
\caption{The distribution of spatial velocity for surviving
companion stars of SNe Ia. The dotted line denotes the results of
ejecta velocity 11000\,km/s, while the solid line shows ejecta
velocity 13500\,km/s.}
\end{figure}

\begin{figure}
\includegraphics[width=5.6cm,angle=270]{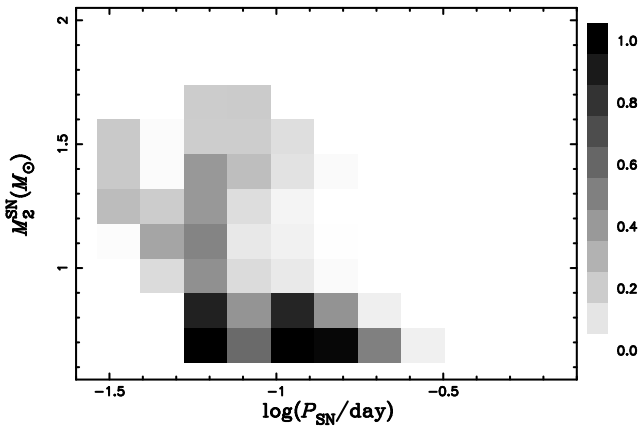}
\caption{Similar to Fig. 1, but in the plane of ($\log P^{\rm SN}$,
$M_2^{\rm SN}$), where $P^{\rm SN}$is the orbital period at the
moment of SN explosion.}
\end{figure}

\begin{figure}
\includegraphics[width=5.6cm,angle=270]{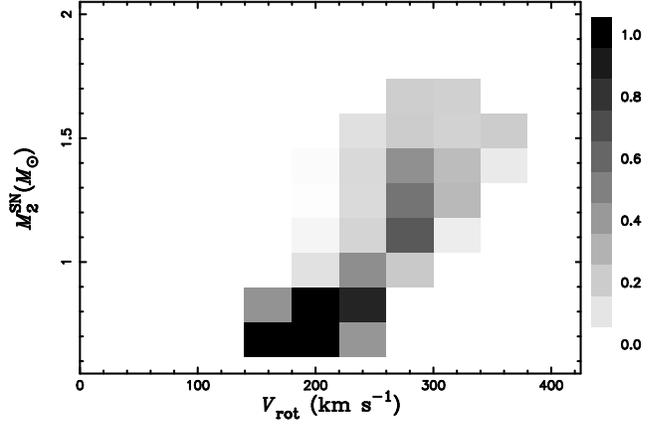}
\caption{Similar to Fig. 1, but in the plane of ($V_{\rm rot}$,
$M_2^{\rm SN}$), where $V_{\rm rot}$ is the equatorial rotational
velocity of companion stars at the moment of SN explosion.}
\end{figure}

Figures 1 and 2 show the distributions of the masses, the orbital
velocities, the effective temperatures, and the surface gravities of
companion stars at the moment of SN explosion. In Figure 1, the
companion star has an orbital velocity of $\sim$300$-$500\,${\rm
km/s}$ for a corresponding mass of $\sim$0.6$-$1.7\,$M_\odot$ at the
moment of SN explosion. In the three formation channels of the WD +
He star systems, SNe Ia mainly come from the He star channel. For
this formation channel, the recorded properties at each step show
that a primordial binary system with a primary mass $M_{\rm 1i}\sim
5.0-8.0\,M_\odot$, a secondary mass $M_{\rm 2i} \sim
2.0-6.5\,M_\odot$, and an orbital period $P_{\rm i} \sim 10-40\,{\rm
d}$ would evolve to a close WD + He star system with a WD mass
$M^{\rm i}_{\rm WD} \sim 0.87-1.2\,M_\odot$, a He star mass $M^{\rm
i}_{\rm 2} \sim 1.0-2.6\,M_\odot$, and an orbital period $P^{\rm i}
\sim 0.04-0.2\,{\rm d}$. Finally, the WD + He star system results in
an SN Ia explosion and survives a companion star.

However, Figures 1 and 2 are for the moment of SN explosion, which
could be modified by the explosion. The SN ejecta will interact with
its companion. The companion stars will be stripped of some mass and
receive a kick velocity that is perpendicular to the orbital
velocity. Adopting a similar method to Meng, Chen \& Han (2007), we
estimated the stripped mass for the companion stars of the WD + He
star channel and find that the stripped mass is very low, e.g., a
1.243$\,M_\odot$ companion star with 0.274$\,R_\odot$ only loses a
mass of 0.015$\,M_\odot$ in our simulation. We also roughly obtain
the kick velocity based on the momentum conservation
equation.\footnote{ However, in reality, the collision between the
SN ejecta and its companion cannot be elastic. If we adopt the
inelastic collision in our calculations, the kick velocity will only
decrease by $\sim$3\%$-$5\%. Thus, the inelastic collision is no
significant influence on the kick velocity.} The kick velocity
mainly depends on the ratio of separation to the radius of
companions at the moment of SN explosion, $A/R_{\rm 2}^{\rm SN}$,
and the leading head velocity of SN ejecta, which is assumed to be
in the range of 11000 to 13500\,km/s. The velocity of 11000\,km/s is
from the SN ejecta kinetic energy $1.0\times10^{51}$\,erg
corresponding to the lower limit of normal SN Ia kinetic energy,
while the velocity of 13500\,km/s is from the SN ejecta kinetic
energy $1.5\times10^{51}$\,erg corresponding to the upper limit of
the kinetic energy (Gamezo et al. 2003). We can obtain the spatial
velocity by the formula $V_{\rm 2}^{\rm SN}=\sqrt{V_{\rm
kick}^{2}+V_{\rm orb}^{2}}$, where $V_{\rm kick}$ and $V_{\rm orb}$
are the kick velocity and the orbital velocity of the companion star
at the moment of SN explosion, respectively. In Figure 3, we show
the distribution of the spatial velocity for the surviving companion
stars of the WD + He star channel. We see that the surviving
companion stars have high spatial velocities ($>$400\,km/s) that
almost entirely exceed the gravitational pull of the Galaxy nearby
the sun. Thus, the surviving companion stars from the WD + He star
channel could be an alternative origin for HVSs.

US 708 is an extremely He-rich sdO star in the Galaxy halo, with a
heliocentric radial velocity of +$708\pm15$\,${\rm km/s}$ (Hirsch et
al. 2005). We note that the local velocity relative to the Galatic
center may lead to a higher observation velocity for the surviving
companion stars, but this may also lead to a lower observation
velocity. Considering the local velocity near the sun
($\sim$220\,km/s), we find that $\sim$30\% of the surviving
companion stars may be observed to have velocity $V>700\,{\rm km/s}$
for a given SN ejecta velocity 13500\,km/s. In addition, the
asymmetric explosion of SNe Ia may also enhance the velocity of the
surviving companions. Thus, a surviving companion star in the WD +
He star channel may have a high velocity like US 708 (see also
Justham et al. 2008).

The companion stars are out of thermal equilibrium at the moment of
SN explosion. For He stars, the equilibrium radii are lower than at
the moment of SN explosion. Thus, the surface gravity at equilibrium
should be greater than the one in Figure 2; e.g., a 1.243$\,M_\odot$
He star with 0.274$\,R_\odot$ and ($\log T_{\rm eff}$, $\log
g$)=(4.70, 5.65) will be a 0.267$\,R_\odot$ He star and with ($\log
T_{\rm eff}$, $\log g$)=(4.75, 5.67) after the He star is back to
the thermal equilibrium. It is well known that a shock will develop
after impact by the ejecta. A large part of the material in the
companion's envelope is heated by the shock, and some of the
material is vaporized from the surface of the companion stars.
Following a similar method to Chen \& Li (2007), we also estimated
the vaporized mass, and find that the mass loss from the companion
stars is not significant ($<$5\%). Such a surviving companion star
may be significantly overluminous or underluminous depending on the
amount of heating (e.g., Podsiadlowski 2003). Figure 2 could be a
starting point for further studies of this kind.

Figure 4 shows the distributions of orbital periods and secondary
masses of the WD + He star systems at the moment of SN explosion.
The orbital periods and secondary masses of the WD + He star systems
at this moment are basic input parameters when one simulates the
interaction between SN ejecta and its companion. It is suggested
that, for hot stars with radiative envelopes (such as He stars),
tidal forces may be inefficient for synchronization (e.g., Zahn
1977). However, the recent study by Charpinet et al. (2008) supports
efficient tidal synchronization for hot subdwarf stars. The work by
Toledano et al. (2007) also indicates, on the other hand, that even
stars with radiative envelopes may have efficient tidal interaction
on a time scale comparable to convective envelopes. Thus, we make an
assumption that the companion stars co-rotate with their orbits. In
Figure 5, we show the distributions of equatorial rotational
velocities of the companions stars. We see that the surviving
companion stars are fast rotators, so their spectral lines should be
broadened noticeably.

The simulation in this {\em Letter} was made with $\alpha_{\rm
ce}\lambda =0.5$. If we adopt a higher value for $\alpha_{\rm
ce}\lambda$, e.g., 1.5, the birthrate of SNe Ia would be a little
bit higher and the delay time from the star formation to SN
explosion longer. This is because binaries emerging from CE
ejections tend to have longer orbital periods for a large
$\alpha_{\rm ce}\lambda$ and are more likely to be located in the SN
Ia production region (Fig. 8 of Wang et al. 2009a). Due to the lack
of WD binaries with short orbital periods ($\log P^{\rm i} < -1.2$)
for a large $\alpha_{\rm ce}\lambda$, the companions with orbital
velocity $V_{\rm orb}>430\,{\rm km/s}$ would be noticeably absent in
Figure 1.

The distributions are results of the current epoch for a constant
SFR. For a single starburst, most of the SN explosions occur between
$\sim$45\,Myr and $\sim$140\,Myr after the starburst; i.e., SNe Ia
from the WD + He star channel will be absent in old galaxies. The
Galactic SN Ia birthrate from this channel is $\sim$$0.3\times
10^{-3}\ {\rm yr}^{-1}$ (Wang et al. 2009b). By multiplying the
birthrate with a typical MS lifetime of He stars, $\sim$$10^7$\,yr,
we estimated the current number of this type of HVSs in the Galaxy
to be $\sim$$10^3$. In future investigations, we will employ the
Large sky Area Multi-Object fiber Spectral Telescope (LAMOST) to
search the HVSs originating from the surviving companion stars of
SNe Ia.

\begin{acknowledgements}
We thank an anonymous referee for his/her valuable comments that
helped us to improve the paper. B.W. thanks Drs S. Justham, X.-C.
Meng and W.-C. Chen for their helpful discussions. This work is
supported by the National Natural Science Foundation of China (Grant
No. 10821061), the National Basic Research Program of China (Grant
No. 2007CB815406), and the Yunnan Natural Science Foundation (Grant
No. 08YJ041001).

\end{acknowledgements}

\clearpage

\end{document}